\documentclass[pre,twocolumn,noshowpacs,a4paper,10pt,nofootinbib]{revtex4-1}

\usepackage{amsmath}
\usepackage{amsfonts}
\usepackage{amssymb}
\usepackage{amsthm}
\usepackage{mathrsfs}
\usepackage{dsfont}
\usepackage{esint}
\usepackage{graphicx}
\usepackage{physics}

\DeclareMathAlphabet{\mathpzc}{OT1}{pzc}{m}{it}

	\newcommand{\brr}[1]{\left[ #1 \right]}

	\newcommand{\off}[1]{\!\brr{#1}}

	\newcommand{\sbrr}[1]{[ #1 ]}

	\newcommand{\soff}[1]{\!\sbrr{#1}}

	\newcommand{\Sum}[2]{\sum\limits_{#1}^{#2}}

	\newcommand{\sSum}[2]{\sum_{#1}^{#2}}

		\newcommand{\Id}{\mathds{1}}

		\newcommand{\sSpan}[1]{\mathrm{Span}\soff{#1}}
		\newcommand{\Span}[1]{\mathrm{Span}\off{#1}}

\newcommand{\Ham}{\hat{H}}
\newcommand{\TEO}{\hat{U}}

\newcommand{\Detect}{\hat{D}}
\newcommand{\PsiDet}{\psi_\text{d}}
\newcommand{\RDet}{r_\text{d}}
\newcommand{\PsiIn}{\psi_\text{in}}
\newcommand{\RIn}{r_\text{in}}
\newcommand{\FDA}{\varphi}
\newcommand{\FDP}{F}

\newcommand{\TDP}{P_\text{det}}

\newcommand{\AUS}{u}
\newcommand{\Stab}{\mathcal{S}_{\Detect}}

\newcommand{\Hilbert}{\mathcal{H}}

\begin{document}

\title{Quantum total detection probability from repeated measurements II. \\ Exploiting symmetry}

\author{Felix Thiel}
\email{thiel@posteo.de}
\author{Itay Mualem}
\author{David A. Kessler}
\author{Eli Barkai}
\affiliation{Department of Physics, Institute of Nanotechnology and Advanced Materials, Bar-Ilan University, Ramat-Gan 52900, Israel}

\begin{abstract}
  A quantum walker on a graph, prepared in the state $\ket{\PsiIn}$, e.g. initially localized at node $\RIn$, 
  is repeatedly probed, with fixed frequency $1/\tau$, to test its presence at some target node 
  $\RDet$ until the first successful detection.
  This is a quantum version of the first-passage problem.
  We investigate the total detection probability $\TDP$, i.e. the probability to eventually detect the 
  particle after an arbitrary number of detection attempts.
  It is demonstrated that this total detection probability is less than unity in symmetric systems,
  where it is possible to find initial states which are shielded from the detector by destructive 
  interference, so-called dark states.
  The identification of physically equivalent initial states yields an upper bound for 
  $\TDP$ in terms of the reciprocal of the number $\nu$ of physically equivalent states.
  The relevant subgroup of the system's symmetry operations is found to be the stabilizer of the detection state.
  Using this, we prove that all bright, i.e. surely detectable, states are symmetric with respect to the stabilizer.
  This implies that $\TDP$ can be obtained from a diagonalization of the ``symmetrized'' Hamiltonian,
  instead of having to find all eigenstates of the Hamiltonian.
\end{abstract}
\maketitle

\section{Introduction}
\label{sec:Intro}
  One of the most fundamental problems in statistical mechanics, with a variety of applications  
  \cite{Redner2007-0, Raposo2009-0, Benichou2011-0, Palyulin2016-0, Godec2016-0, Godec2016-1}, is the 
  first-passage problem: finding the distribution of the first time that a random walker, initially 
  in position $\RIn$, will reach its destination $\RDet$.
  The quantum first detection problem \cite{ Bach2004-0, Krovi2006-0, Krovi2006-1, Krovi2007-0, %
    Stefanak2008-0, Varbanov2008-0, Caruso2009-0, Agliari2010-0, Gruenbaum2013-0, Bourgain2014-0, %
    Krapivsky2014-0, Dhar2015-0, Dhar2015-1, Sinkovicz2015-0, Sinkovicz2016-0, Lahiri2019-0, %
    Friedman2017-0, Friedman2017-1, Thiel2018-0, Thiel2018-1%
  } is a quantum version of the first-passage problem.
  Here a quantum system is prepared in some state $\ket{\PsiIn}$, e.g. a localized state on a graph $\ket{\RIn}$,
  and one seeks the first time that it is {\em detected} in the target state $\ket{\PsiDet}$, which may also be a localized state $\ket{\RDet}$.
  The conceptual complications related to quantum trajectories, observation and wave-function collapse are overcome by adhering to a 
  detection protocol that combines unitary evolution with repeated detection attempts every $\tau$ time units.
  In each detection event, the observer tries to detect the system in $\ket{\PsiDet}$ until he is successful 
  for the first time.
  This stroboscopic detection protocol defines the time of first {\em detected} arrival to $\ket{\PsiDet}$.
  The procedure is closely related to the time-of-arrival problem \cite{Allcock1969-0, Kijowski1974-0, %
    Aharonov1998-0, Damborenea2002-0, Anastopoulos2006-0, Halliwell2009-0, Sombillo2014-0, Sombillo2016-0%
  }, and to conventional quantum search setups \cite{Grover1997-0, Aaronson2003-0, Bach2004-0, %
    Childs2004-0, Muelken2006-0, Perets2008-0, Karski2009-0, Zaehringer2010-0, Muelken2011-0, Jackson2012-0,%
    Novo2015-0, Boettcher2015-0, Preiss2015-0, Xue2015-0, Li2017-0, Mukherjee2018-0, Rose2018-0%
  }.

  In this series of articles \cite{Thiel2019-0,Thiel2019-2}, we investigate the total detection probability $\TDP$.
  This is the probability that the quantum particle is at all detected under the stroboscopic detection protocol,
  which -- in contrast to classical ergodic random walks -- is in general not unity, not even on finite structures.
  In the first article \cite{Thiel2019-0} (hereafter referred to as [I]), 
  we discussed the dark and bright states, special initial conditions that will never or surely lead to detection, respectively.
  The identification of all dark and bright energy eigenstates led to an explicit formula for $\TDP$ in terms of
  the Hamiltonian's spectral decomposition.
  The dark states that are responsible for the deficit in $\TDP$ arise either from energy levels that 
  have no overlap with the detection state or from degenerate energy levels.

\begin{figure}
    \centering
    \includegraphics[width=0.99\columnwidth]{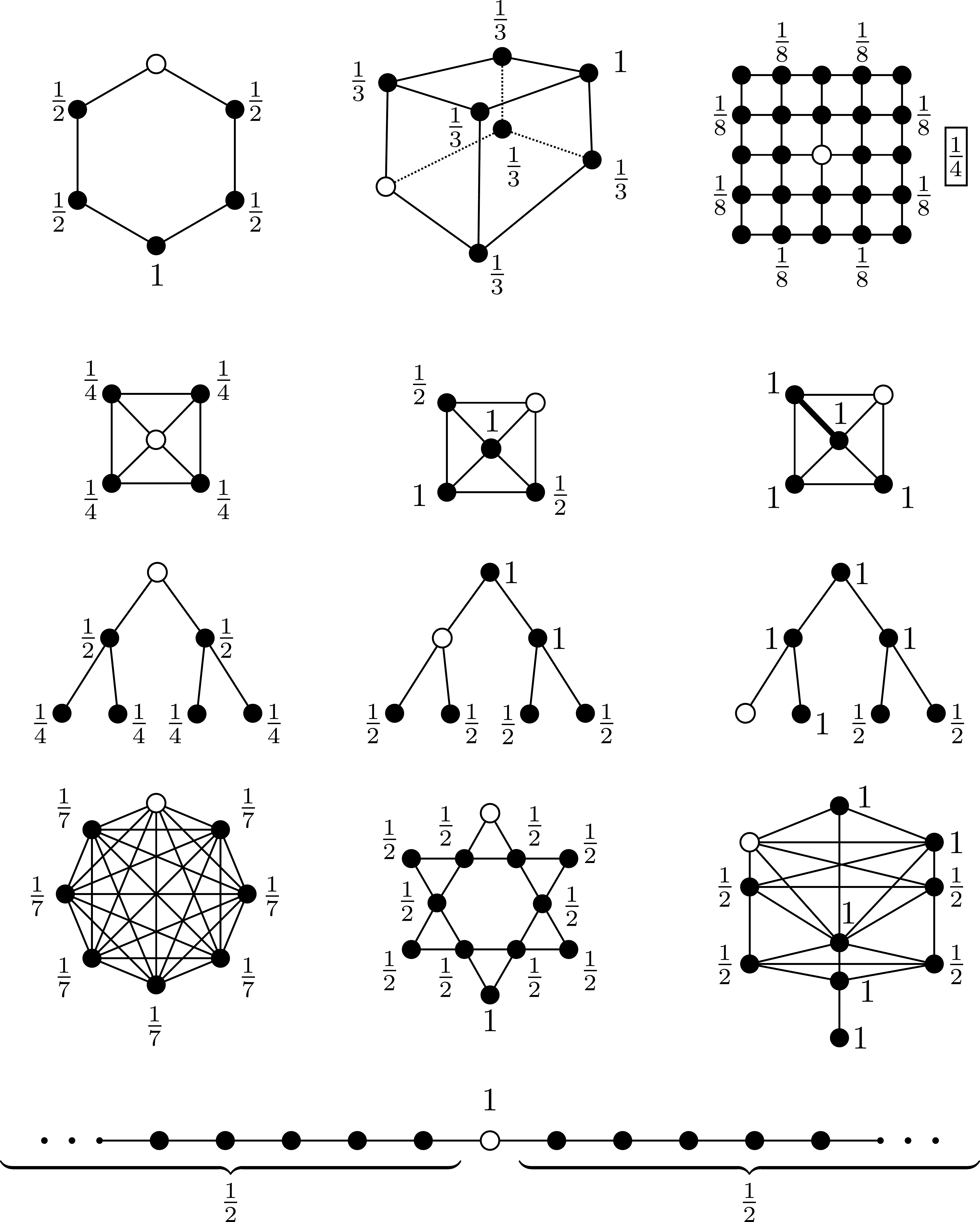}
    \caption{
      The upper bound for some simple graphs.
      The numbers represent the bound for $\TDP$ from Eq.~\eqref{eq:UpperBound}.
      An open circle denotes the detection site $\RDet$, and any other node is a possible localized initial state $\ket{\PsiIn} = \ket{\RIn}$.
      The quantum particle resides on the nodes of these graphs and travels along its links.
      In all graphs, the on-site energies are equal to zero.
      From left to right and top to bottom:
      The ring of size six, the hypercube of dimension three, a two dimensional simple cubic lattice, 
      the square graph with detection site in the center, in a corner, and in a corner with with one modified link,
      the binary tree graph in two generations with detection in the root, middle and leaves,
      the complete graph with eight sites, the Star-of-David graph, and the Tree-of-Life graph.
      The infinite line is shown at the very bottom.
      The numbers are upper bounds for $\TDP$ and some of them may be compared with the exact results 
      obtained in [I].
    }
    \label{fig:PD}
  \end{figure}

  In this article, we continue our investigation of $\TDP$, exploring the consequences of the system's symmetries.
  We exploit the connection between symmetry and degeneracy \cite{Weyl1950-0, Wigner1959-0}
  and explore what can be inferred about the total detection probability from symmetry grounds alone.
  In particular, we present a simple upper bound that quantifies $\TDP$ in terms of the number of
  ``physically equivalent'' states, see below and see Fig.~\ref{fig:PD}, where we have summarized a set of examples.
  In many cases, this upper bound is saturated, and $\TDP$ is directly determined.
  We discuss under which conditions this is the case.

  The rest of the paper is organized as follows:
  We recapitulate our model, the main quantities and relevant findings of [I] in sec.~\ref{sec:Prelim}.
  A simple argument explaining why one expects to find dark states in symmetric systems follows in sec.~\ref{sec:WhyDarkStates}.
  Our main findings are presented in sec.~\ref{sec:Symmetry}, where also the examples are discussed.
  Sec.~\ref{sec:Stabilizer} is devoted to formalizing these results and it explains under which 
  conditions the upper bound becomes an equality.
  Based on this, we outline a dimensionality reduction of the original problem in sec.~\ref{sec:SymHam},
  and close with a summary and discussion in sec.~\ref{sec:Summary}.
  Some technical complications are relegated to the appendix.

\section{Preliminaries}
\label{sec:Prelim}
  We consider the quantum dynamics of a time-independent Hamiltonian $\Ham$.
  The measurement-free unitary dynamics are governed by the Schr\"odinger equation which yields the evolution operator $\TEO(t) = e^{-i t \Ham / \hbar}$.
  The system is initially in the state $\ket{\PsiIn}$ and each $\tau$ time units, a strong measurement with the 
  projector $\Detect := \dyad{\PsiDet}$ probes whether it is in the detection state $\ket{\PsiDet}$ or not.
  In case the detection was unsuccessful, the wave function is wiped of its $\ket{\PsiDet}$ component, then renormalized, 
  and is used in the next evolution / attempted detection double-step.
  This is repeated until the first successful detection at, say, the $n$-th attempt.
  The time $T = n \tau$ is the quantum first detection time.
  The stroboscopic detection protocol used herein serves as $T$'s operational definition; for more information see \cite{Friedman2017-1, Thiel2019-0}.
  The probability of the event $T = n\tau$ is the {\em first detection probability} $F_n(\PsiIn) = \abs*{\FDA_n(\PsiIn)}^2$,
  which can be written as the squared modulus of the {\em first detection amplitudes} given by \cite{Dhar2015-0, Dhar2015-1, Friedman2017-1}:
  \begin{equation}
    \FDA_n(\PsiIn)
    = 
    \mel*{\PsiDet}{\TEO(\tau)[(\Id-\Detect) \TEO(\tau)]^{n-1}}{\PsiIn}
    .
  \label{eq:DefFDA}
  \end{equation}
  Reading the equation right-to-left, we see that the initial state 
  is subject to $n-1$ double steps of unitary evolution and unsuccessful detection, 
  until finally, after the $n$-th evolution step, detection is successful.
  The main focus of this paper lies in the {\em total detection probability}:
  \begin{equation}
    \TDP(\PsiIn)
    =
    \Sum{n=1}{\infty} F_n(\PsiIn)
    = 
    \Sum{n=1}{\infty} \abs{\FDA_n(\PsiIn)}^2
    ,
  \end{equation}
  which is the probability to detect the particle at all.
  $F_n := \abs{\FDA_n}^2$ is the probability of first detection at the $n$-th attempt.
  Our notation stresses the dependence on the initial state.
  The dependence on the detection state, however, will be suppressed throughout the article.

  In [I], we showed among other things that the detection protocol splits a finite-dimensional Hilbert space 
  into a bright and a dark part $\Hilbert = \Hilbert_B \oplus\Hilbert_D$.
  Each of these subspaces is invariant under the detection protocol.
  Each bright state $\ket*{\widetilde{\beta}} \in \Hilbert_B$ is detected with probability one, $\TDP(\widetilde{\beta}) =1$, 
  and no dark state $\ket*{\widetilde{\delta}} \in \Hilbert_D$ can ever be detected, $\FDA_n(\widetilde{\delta}) = 0$.
  Every initial state $\ket{\PsiIn}$ is either bright or dark or a superposition of a bright and a dark state.
  As its dark component does not contribute to $\TDP$, the total detection probability of $\ket{\PsiIn}$ must be equal to 
  the initial state's overlap with the bright space \cite{Krovi2006-0}.
  
  We assumed in [I] that a diagonalization in terms of discrete eigenphases of the evolution operator is available:
  \begin{equation}
    \TEO(\tau)
    =
    \Sum{l}{} e^{- i \lambda_l} \hat{P}_l
    \qc
    \hat{P}_l := \Sum{m=1}{g_l} \dyad{E_{l,m}}
    ,
  \label{eq:TEODiag}
  \end{equation}
  where $\ket{E_{l,m}}$ are the eigenstates, $\hat{P}_l$ are the eigenspace projectors,
  $g_l$ are the degeneracies and $\lambda_l := E_l\tau / \hbar \mod 2\pi$ are the distinct phases that appear in the evolution operator,
  and are derived from the energy levels $E_l$ of the Hamiltonian modulo $2\pi \hbar/\tau$.
  There is a one-to-one correspondence between the energy levels $E_l$ and the phases $\lambda_l$,
  as long as resonant detection periods $\tau_c$, defined by 
  \begin{equation}
    \frac{\tau_c}{\hbar}
    \abs{E_l - E_{l'}} 
    = 0 \mod 2\pi
    ,
  \label{eq:DefResonantTau}
  \end{equation}
  for some pair of energy levels, are avoided.
  Each $\lambda_l$-sector that has overlap with the general detection state $\ket{\PsiDet}$ 
  contains exactly one bright eigenstate, i.e. whenever $\hat{P}_l\ket{\PsiDet} \ne 0$, 
  we defined the bright eigenstates
  \begin{equation}
    \ket{\beta_l} 
    := 
    \frac{\hat{P}_l\ket{\PsiDet}}{\sqrt{\smash[b]{\ev*{\hat{P}_l}{\PsiDet}}}}
    .
  \label{eq:DefBrightEigen}
  \end{equation}
  The $\ket{\beta_l}$ are a basis for $\Hilbert_B$.
  Summing the squared overlaps of $\ket{\PsiIn}$ with each $\ket{\beta_l}$, we arrived at our main formula:
  \begin{equation}
    \TDP(\PsiIn)
    =
    \sideset{}{'} \sum_l 
    \frac{
      \abs*{
        \sSum{m=1}{g_l} \ip{\PsiDet}{E_{l,m}} \ip{E_{l,m}}{\PsiIn} 
      }^2
    }{
      \sSum{m=1}{g_l} \abs*{\ip{\PsiDet}{E_{l,m}}}^2
    }
    ,
  \label{eq:Kessler}
  \end{equation}
  where the sum excludes all $l$ for which the denominator would vanish.
  Note that this formula requires the knowledge of {\em all} $\lambda_l$ and $\ket{E_{l,m}}$,
  i.e. a full diagonalization of the evolution operator.
  An important feature of this result is the apparent lack of dependence on $\tau$.
  In fact $\TDP(\PsiIn)$ does not depend on $\tau$ except for the resonant values $\tau_c$ defined above.
  At these points, the number of quasienergy levels jumps and with it their degeneracies.
  For details, we refer the reader to [I].

  Our theory is developed in generality, but our examples are taken exclusively from the so-called continuous-time quantum walks \cite{Muelken2011-0}.
  These describe single quantum particles residing on the nodes of a finite graph, propagating along its edges.
  Their Hamiltonian $\Ham$ is the product of the graph's adjacency matrix \cite{Bapat2014-0} and an energy constant $\gamma$.
  The natural basis of the Hilbert space consists of states $\ket{r}$ localized on the nodes $r$ of the graph.

  Throughout the main part of this text, we make the following assumptions: 
  that the detection state is localized $\ket{\PsiDet} = \ket{\RDet}$ and that the resonant detection 
  periods of Eq.~\eqref{eq:DefResonantTau} are avoided.
  These assumptions are not essential, but rather allow for a simpler presentation, skirting some
  technicalities that are treated in appendix~\ref{app:Peculiarities}.
  Additionally, we assume in sec.~\ref{sec:Symmetry} that the initial state is localized, 
  which is immediately generalized in the following section.
  We use the symbol $\ket{\psi}$ for general states in the Hilbert space and 
  $\ket{r}$ for states that are localized in the position (graph node) basis.

  Next, we give a short argument why one expects to find dark states.

\section{Why do we find dark states?}
\label{sec:WhyDarkStates}
  In the absence of measurements, the wave function is 
  \begin{equation}
  \ket{\psi_{{\rm free}}(t)} = \TEO(t)\ket{\PsiIn}= e^{- i t \frac{\Ham}{\hbar}}  \ket{\PsiIn}
  \label{eq02}
  \end{equation} 
  where the subscript ``free'' means an evolution free of a measurement process. 
  If 
  \begin{equation}
    \ip*{\RDet}{\psi_\text{free}(t)} 
    =
    \mel*{\RDet}{\TEO(t)}{\PsiIn} 
    = 0
    \qc \text{for any } t
  \label{eq:DefGenDark}
  \end{equation} 
  then the target state cannot be detected, since the probability of successful detection is zero at all times.
  Also, the unsuccessful measurements have no effect on the system.
  Hence the state $\ket{\PsiIn}$ is dark.

  Consider now the situation, when there are two different initial states $\ket{\PsiIn}$ and 
  $\ket{\PsiIn'}$ which yield identical transition amplitudes to the detection state:\footnote{
    Later, in Appendix~\ref{app:Peculiarities}, we will relax this condition in that we only 
    require identical transition {\em probabilities}.
    The {\em amplitudes} in left and right-hand side of Eq.~\eqref{eq:DefPhysEq} may thus 
    differ by an irrelevant phase factor $e^{i\lambda}$.
  }
  \begin{equation}
    \mel*{\RDet}{\TEO(t)}{\PsiIn} 
    =
    \mel*{\RDet}{\TEO(t)}{\PsiIn'}
    \ne 0
    .
  \label{eq:DefPhysEq}
  \end{equation}
  We will call such a pair of states {\em physically equivalent to each other}.
  Clearly, these can only appear in systems with a certain degree of symmetry.
  From these two we can construct the normalized initial superposition state $(\ket{\PsiIn} - \ket{\PsiIn'})/\sqrt{2}$,
  but this must be a dark state according to Eqs.~(\ref{eq:DefGenDark},~\ref{eq:DefPhysEq}).
  Hence, any pair of physically equivalent states yields a dark state.
  This reveals dark states as an interference phenomenon.
  Certain initial states result in permanent destructive interference and thus vanishing probability amplitude in the detection state.

\section{From symmetries to the detection probability} 
\label{sec:Symmetry}
  In this section we assume that the initial state is localized:
  \begin{equation}
    \ket{\PsiIn} = \ket{\RIn}
    .
  \label{eq:}
  \end{equation}
  This way, some of our definitions can be understood more intuitively.
  The generalization for non-localized initial states follows in the next section.

  From Eq.~\eqref{eq:DefFDA}, we see that the detection amplitude $\FDA_n$ is linear with respect 
  to the initial state, and thus obeys a superposition principle.
  This allows us to obtain an upper bound for $\TDP(\RIn)$ without the need of detailed calculations. 
  To see this, consider localized initial and detection states $\ket{\RIn}$ and $\ket{\RDet}$. 
  Now we introduce an auxiliary initial state which is a linear combination
  of any two different localized states
  \begin{equation}
    \ket{\AUS_\alpha} 
    = 
    \frac{1}{\sqrt{2}}
    ( \ket{\RIn} + e^{i\alpha} \ket{r'} )
    ,
  \label{eq:SuperPosInitial}
  \end{equation}
  where $\ip{\RIn}{r'} = 0$ and $\alpha$ is an arbitrary relative phase.
  As the first detection amplitudes are linear in the initial state, see Eq.~\eqref{eq:DefFDA}, we find
  \begin{equation}
    \FDA_n(\AUS_\alpha) 
    =
    \frac{1}{\sqrt{2}} \qty[
      \FDA_n(\RIn)
      + e^{i\alpha} \FDA_n(r')
    ]
    .
  \label{eqRE10}
  \end{equation}

  We can use this to find a very useful upper bound on the detection probability,
  provided there is a symmetry relation between $\RIn$ and $r'$, namely, when $\ket{\RIn}$ and $\ket{r'}$ 
  are physically equivalent, see Eq.~\eqref{eq:DefPhysEq}.
  For example in the ring system of Fig.~\ref{fig:PD}, the two sites left and right of the detection node 
  are equivalent due to reflection invariance.
  Clearly, when $\ket{\RIn}$ and $\ket{r'}$ are physically equivalent then also $\FDA_n(\RIn) = \FDA_n(r')$.
  Under such circumstances, for the superposition $\ket{\AUS_\alpha}$ in Eq.~\eqref{eq:SuperPosInitial}, we have:
  \begin{equation}
    \FDA_n(\AUS_\alpha)
    =
    \frac{1 + e^{i\alpha}}{\sqrt{2}}
    \FDA_n(\RIn)
  \label{eqRE11}
  \end{equation}
  and so
  \begin{equation}
    F_n(\AUS_\alpha)
    =
    \abs{ \FDA_n(\AUS_\alpha) }^2
    =
    ( 1 + \cos\alpha ) F_n(\RIn)
  .
  \label{eqRE11}
  \end{equation}
  The relative phase $\alpha$ affects the first detection statistics.
  The particular choice $\alpha=\pi$ is the situation considered in sec.~\ref{sec:WhyDarkStates} and yields a dark state.
  A useful bound on $\TDP$ can be found by summing over all $n$ and using $\TDP(\AUS_\alpha) = \sSum{n=1}{\infty} F_n(\AUS_\alpha) \le 1$:
  \begin{equation}
    1 \ge 
    \TDP(\AUS_\alpha)
    = 
    (1 + \cos\alpha) \TDP(\RIn)
    .
  \label{eqRE12}
  \end{equation}
  Choosing $\alpha=0$, we obtain for the originally considered transition from $\RIn$ to $\RDet$:
  \begin{equation}
    \TDP(\RIn)
    \le 
    \frac{1}{2}
    .
  \label{eqRE13}
  \end{equation}
  Thus, if $\RIn$ has a physically equivalent partner, the total detection probability cannot be unity, 
  unlike for the classical random walk.

  \begin{figure}
    \centering
    \includegraphics[width=0.99\columnwidth]{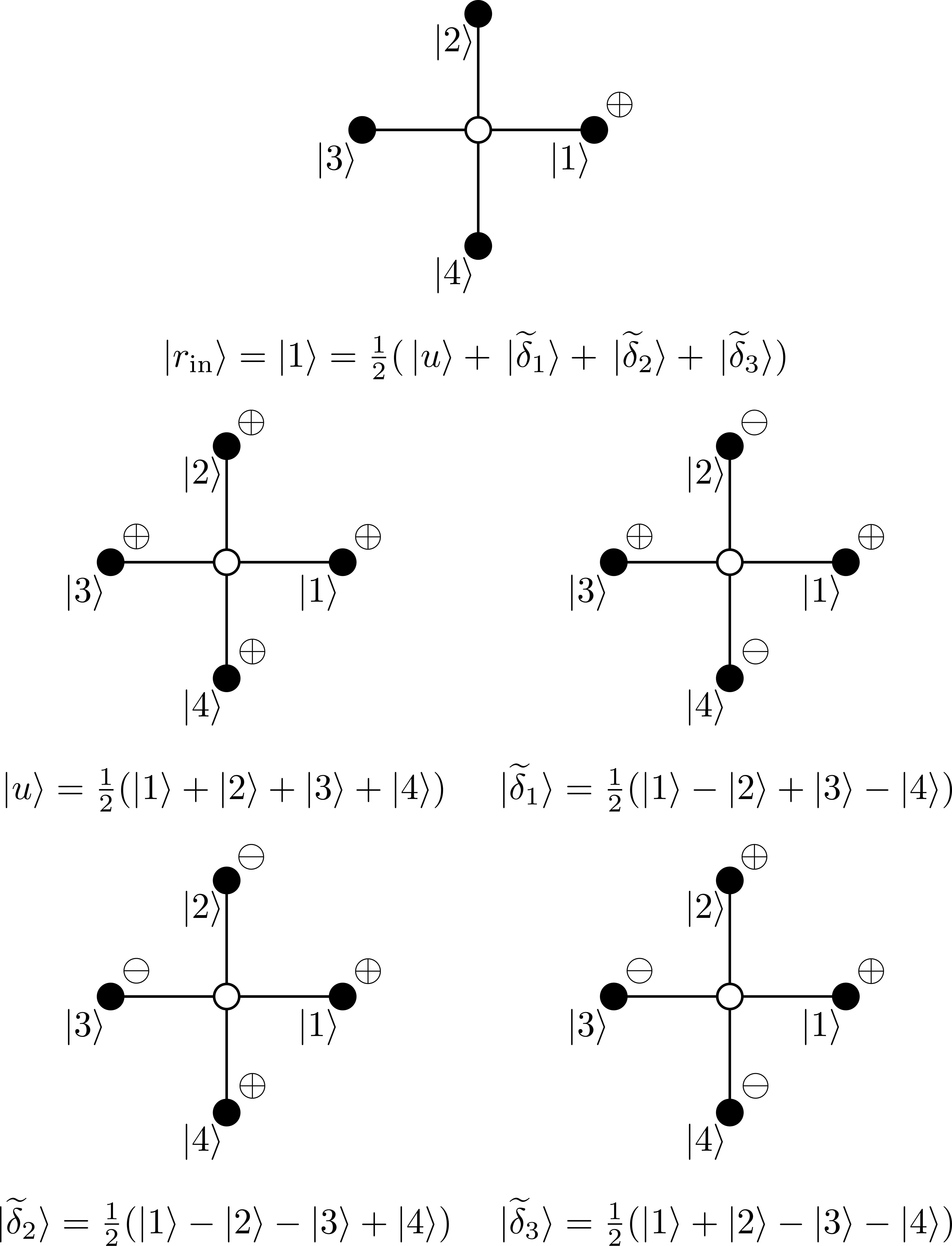}
    \caption{
      Decomposition of an initial state $\ket{1}$, localized on an exterior node of a cross, into three dark components,
      and the auxiliary uniform  state (AUS).
      Detection is attempted on the node in the center (open circle).
      We find $\nu = 4$ from the $\pi/2$ degree rotation symmetry about the detection site.
      Consequently, $\TDP(1) \le  1/4$ from the simple bound, Eq.~\eqref{eq:UpperBound}. 
    }
    \label{fig3F}
  \end{figure}

  This bound can be easily generalized to other structures, for example the cube of Fig.~\ref{fig:PD}.
  Consider the transition to one vertex, denoted by $\ket{0}$, from one of its nearest neighbors, say $\ket{1}$. 
  Let us denote the neighbors of $\ket{0}$ by $\ket{1}$, $\ket{2}$, and $\ket{3}$ and define an 
  auxiliary state $\ket{u} := (\ket{1} + \ket{2} + \ket{3})/\sqrt{3}$.
  (This would be the generalization of $\ket{\AUS_0}$ from before.)
  Using the same procedure as before, we find that $\TDP(1) \le 1/3$.
  This trick can be easily extended and is summarized in the following proposition:

  {\em Proposition 1:} 
  The total detection probability for the transition from a localized initial state $\ket{\RIn}$ to 
  a localized detection state $\ket{\RDet}$ is bounded by the reciprocal of the number $\nu$ of 
  nodes physically equivalent to $\RIn$:
  \begin{equation}
    \TDP(\RIn)
    \le 
    \frac{1}{\nu}
    .
  \label{eq:UpperBound}
  \end{equation}

  Recall the definition of physically equivalent states from Eq.~\eqref{eq:DefPhysEq}:
  They have identical transition amplitudes to the detection state at all times.
  Such states are indistinguishable in the first detection problem.
  For a given system and a given transition $\RIn\to\RDet$ from one localized detection state to another, 
  we can find a set of $\nu$ initial nodes $\{ r_j \}_{j=0}^{\nu-1}$ (where $r_0 = \RIn$) which are
  physically equivalent in this sense.
  They can be identified from elementary symmetry considerations.
  This number $\nu$ of physically equivalent states is what appears in Eq.~\eqref{eq:UpperBound}.
  The strongest bound is given by the {\em maximal} number of physically equivalent states,
  i.e. by the largest possible value for $\nu$.
  However, Eq.~\eqref{eq:UpperBound} still holds even when this maximal number can not be identified beyond doubt.

  Let's discuss the remaining examples of Fig.~\ref{fig:PD}.
  For the cube, we find $\TDP \le 1/3$ except for the diametrically opposed node, for which $\TDP \le 1$.
  For the complete graph with $L$ sites, all nodes besides the detection site are equivalent 
  and we find $\TDP \le  1/(L-1)$ (here $L=8$).
  For simple-cubic lattices in $d$ dimensions with periodic boundary conditions one finds $\TDP \le 1/(2d)$ 
  for sites on the main horizontal, vertical and on the main diagonals, but smaller values for sites which are off these main axes.
  The Star-of-David graph with detector on one of the tips yields $\TDP \le 1/2$, due to reflection symmetry, except for the opposing tip.
  Similarly, yet less obvious $\TDP \le 1/2$ in the Tree-of-Life graph.
  For a square graph with an additional node in the center, we have $\nu=4$ for a transition from one of the 
  corners to the center.
  When the detector is in one of the square's corners, the neighboring corners are equivalent yielding $\nu=2$,
  but all other sites are unique.
  Also for a tree, the number of physical equivalent sites and thus the upper bound varies with 
  the position of the detector.
  The square, the ring, the complete graph, the hypercube, and the tree have been discussed 
  in [I], where we computed $\TDP$ exactly.
  With one exception, the exact values of these examples actually coincide with the upper bound!
  The only exception is the tree, when the detector is not placed on the root node.
  This example will be discussed later.

  Finally, the infinite line demands special attention.
  The upper bound yields $\TDP \le 1/2$ for every non-detection site and is correct.
  However, since it is an infinite system, the theory of [I] does not strictly apply.
  In particular one finds that $\TDP$ has a complicated dependence on $\tau$, see Refs.~\cite{Friedman2017-1, Thiel2018-0},
  where this model was investigated in detail.
  Yet, the $\TDP(\tau)$ curves stay below the upper bound $1/2$.

  To gain further physical insight, consider the cross structure presented in Fig.~\ref{fig3F}. 
  We detect on the center of the cross, at node $\ket{0}$ and start on one of the outer nodes, for example on state $\ket{1}$.
  This initial state can be decomposed into a linear combination of four states, out of which three are
  easily understood as being dark states. 
  For example the  state $\ket*{\tilde{\delta_1}}=(\ket{1} -\ket{2} + \ket{3} - \ket{4})/2$ is dark 
  since it is not injecting probability current into state $\ket{0}$. 
  Destructive interference erases all amplitude in the detection state.
  The uniform state $\ket{\AUS} =(\ket{1} + \ket{2} + \ket{3} + \ket{4})/2$ 
  which is a normalized sum of all the equivalent states in the system, is the fourth state.
  This state gives a constructive interference pattern at the detected state.
  Returning to the transition $\ket{1} \to \ket{0}$ we decompose the initial condition into a 
  superposition of the four states, as shown in Fig. \ref{fig3F}. 
  Since three components are dark, and the overlap of initial and uniform state is $1/\sqrt{4}$ we find $\TDP(1) \le 1/4$.
  In fact, a straight-forward calculation using Eq.~\eqref{eq:Kessler} shows $\TDP(u) = 1$ and thus $\TDP(1) = 1/4$. 
  This is clearly in accord with $\nu = 4$.

  Let us formalize our result.
  Consider a localized initial state $\ket{\RIn} $ and assume that in
  the system we have a total of $\nu$ physically equivalent states $\{ \ket{r_j} \}_{j=0}^{\nu-1}$,
  where $\ket{r_0} = \ket{\RIn}$ and $\ip{r_i}{r_j} = \delta_{i,j}$.
  These states span the subspace $\mathcal{E}_{\ket{\RIn}} = \Span{\ket{r_j}}$, i.e. the space of all possible
  superpositions of the physically equivalent states $\ket{r_j}$.
  Let 
  \begin{equation}
    \ket*{\AUS(\RIn)}
    :=
    \frac{1}{\sqrt{\nu}} \Sum{j=0}{\nu-1} \ket{r_j}
  \label{eq:DefAUS}
  \end{equation}
  be the {\em auxiliary uniform state} (AUS).
  $\ket{\AUS(\RIn)}$ is one particular state in the subspace $\mathcal{E}_{\ket{\RIn}}$.
  We can find a new basis $\{ \ket{\AUS(\RIn)}, \ket*{\widetilde{\delta}_1}, \hdots, \ket*{\widetilde{\delta}_{\nu-1}} \}$
  for $\mathcal{E}_{\ket{\RIn}}$, that consists of the AUS and $\nu-1$ dark states.
  These dark states can be found similarly to the stationary dark states in [I].
  The solution reads:
  \begin{equation}
    \ket*{\widetilde{\delta}_j}
    =
    \frac{j \ket{r_{j}} - \sSum{m=0}{j-1} \ket{r_m}}{\sqrt{j (j+1) } }
    ,
  \label{eq:DarkPhysEqSol}
  \end{equation}
  where $j=1,2,\hdots, \nu-1$.
  One quickly verifies that the states $\ket*{\widetilde{\delta}_j}$ are normalized, orthogonal to each other and to the AUS.
  Furthermore they are dark, because the transition amplitudes to the detection state from any state $\ket{r_j}$ are the same.
  Consequently, the multiplication with $\bra{\RDet}\TEO(t)$ annihilates $\ket*{\widetilde{\delta}_j}$, because it yields
  $\mel*{\RDet}{\TEO(t)}{\widetilde{\delta}_j} = \mel*{\RDet}{\TEO(t)}{\RIn} [ j - j ]/ \sqrt{j(j+1)} = 0$.
  The original initial state $\ket{\RIn} = \ket{r_0}$ is now rewritten as 
  \begin{equation}
    \ket{\RIn}
    =
    \frac{1}{\sqrt{\nu}} \ket*{\AUS(\RIn)}
    - \Sum{j=1}{\nu-1} \frac{\ket*{\widetilde{\delta}_j}}{\sqrt{j(j+1)}} 
    .
  \label{eq:}
  \end{equation}
  Knowing about the dark states in $\ket{\RIn}$, we can directly obtain $\TDP(\RIn)$, 
  because $\FDA_n(\widetilde{\delta}_j) = 0$.
  We immediately find:
  \begin{equation}
    \TDP(\RIn)
    =
    \frac{1}{\nu} \TDP(u(\RIn))
    .
  \label{eq:AUSAndNu}
  \end{equation}
  Since $\TDP(\AUS) \le 1$, we obtain the upper bound Eq.~\eqref{eq:UpperBound}.
  It is important to keep in mind that $\nu$ is not a global property of the system,
  but depends on the transition $\RIn\to\RDet$.
  Clearly, if the AUS is bright, then $\TDP(\AUS(\RIn)) = 1$, and the upper bound saturates,
  i.e. it becomes an equality.
  We also note, that Eq.~\eqref{eq:AUSAndNu} holds as well in infinite systems.
  However, the determination of $\nu$ can become non-trivial in this situation.

\section{Symmetry properties of the bright states}
\label{sec:Stabilizer}
  \begin{figure}
    \includegraphics[width=0.99\columnwidth]{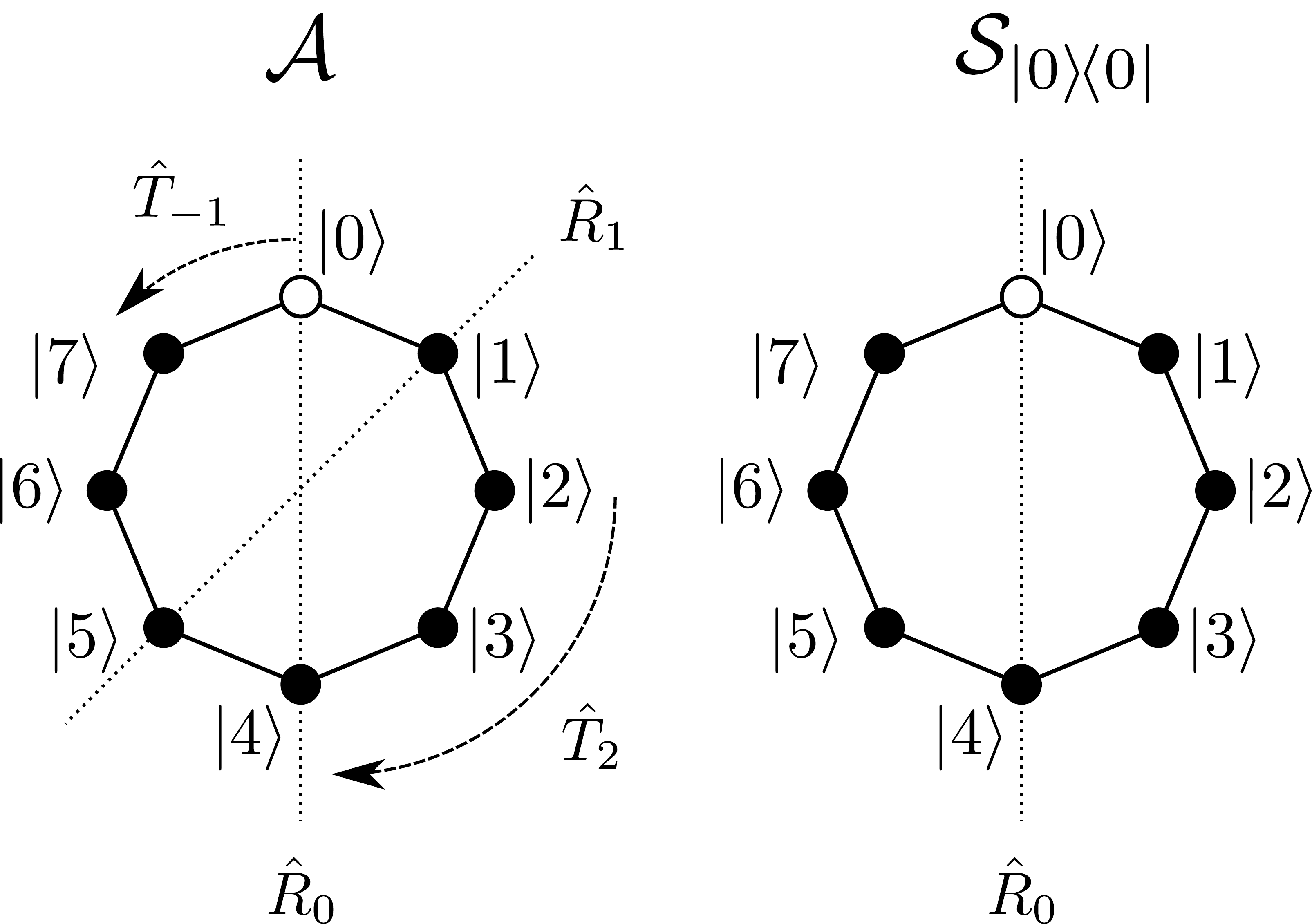}
    \caption{
      The Schr\"odinger group $\mathcal{A}$ of a ring with $8$ sites and the stabilizer subgroup 
      $\mathcal{S}_{\dyad{0}}$ of the detection site.
      Left: The symmetry group consists of all translations and reflections of which we sketched two each.
      Namely the shift by two sites to the right, the shift by one site to the left and the reflections
      around site one and site zero.
      Right: Detection takes place in site $\ket{0}$.
      The stabilizer subgroup $\mathcal{S}_{\dyad{0}}$ consists of all those symmetry transformations
      that do not change the detection state.
      Here this is only the reflection $\hat{R}_0$ around the detection site itself and the identity.
      The stabilizer subgroup represent the relevant symmetries that determine which states are 
      physically equivalent.
      In the ring we find $\nu = 2$ for all sites, except for the detection site and the site 
      opposing the detection site, which have $\nu = 1$.
      \label{fig:SymmetryTransforms}
    }
  \end{figure}
  In [I], we have expressed the total detection probability in terms of the 
  initial state's overlap with the bright space.
  We have found the bright eigenstates, Eq.~\eqref{eq:DefBrightEigen}, and used them in
  Eq.~\eqref{eq:Kessler} to find $\TDP(\PsiIn)$.
  In Eq.~\eqref{eq:AUSAndNu}, however, we have expressed this quantity in terms of the number of 
  physically equivalent states and the AUS.
  In this section we will connect these two approaches by refining our definition of physically 
  equivalent states and of what ``symmetry around the detection state'' means.
  This will reveal when the upper bound saturates and why -- as mentioned -- many of 
  the upper bounds in Fig.~\ref{fig:PD} coincide with the exact results of [I].
  Although we still assume that the detection state $\ket{\RDet}$ is localized, we no longer 
  require the initial state $\ket{\PsiIn}$ to be localized.
  
  What are the system's symmetries?
  A symmetry operation is represented by an operator $\hat{A}$.
  This operator must be unitary, so that it does not change the norm of a state.
  Furthermore it must not change the system's dynamics, that means it must commute with 
  the evolution operator $\TEO(\tau)$ of Eq.~\eqref{eq:TEODiag}.
  When the detection period is not tuned to one of the resonant values of Eq.~\eqref{eq:DefResonantTau},
  there is a one-to-one correspondence between the spectrum of $\TEO(\tau)$ and the Hamiltonian's.
  Therefore, we can also assume that $\hat{A}$ commutes with $\Ham$.
  The set $\mathcal{A}$ of all such operators is called the system's symmetry or {\em Schr\"odinger group}:\footnote{
    In the case of a quantum walk on a graph, when the Hamiltonian is equal to the adjacency
    matrix of the graph, the Schr\"odinger group is equal to the graph's automorphism group.
    The symmetries are often directly visible from the graph.
  }
  \begin{equation}
    \mathcal{A}
    :=
    \{ \hat{A} \text{ is unitary} \, | \, \comm*{\hat{A}}{\Ham} = 0 \}
    .
  \label{eq:DefSchroedingerGroup}
  \end{equation}
  As a group it satisfies closedness and associativity of the group operation, existence of the 
  identity, and existence of inverse elements \cite{Weyl1950-0, Wigner1959-0}.

  Throughout this section, we will treat the Hamiltonian on a ring with $L$ sites as an example:
  \begin{equation}
    \Ham 
    =
    - \gamma \Sum{r=0}{L-1} \qty[ \dyad{x}{x+1} + \dyad{x}{x-1} ]
  \label{eq:TBHam}
  \end{equation}  
  where we identify $\ket{L+r} = \ket{r}$ and $\ket{-r} = \ket{L-r}$.
  The Schr\"odinger group $\mathcal{A}$ for this system consists of all translations $\hat{T}_\xi$ 
  by $\xi$ sites and all reflections $\hat{R}_r$ around site $r$, see the left hand side of Fig.~\ref{fig:SymmetryTransforms}.
  \begin{align}
    \hat{T}_\xi
    :=
    \Sum{r=0}{L-1}
    \dyad{r+\xi}{r}
    \qc
    \hat{R}_{r}
    :=
    \Sum{\xi=0}{L-1}
    \dyad{r-\xi}{r+\xi}
    .
  \label{eq:DefSymmOpRing}
  \end{align}
  As is well known \cite{Weyl1950-0, Wigner1959-0}, non-trivial, i.e. non-commuting, symmetries in a system always imply a 
  degeneracy of energy levels.
  This degeneracy -- as we have shown in [I] -- leads to dark states.
  In the case of the ring, translations and reflections do not commute;
  we have $\hat{T}_\xi \hat{R}_r \hat{T}_{-\xi} = \hat{R}_{r+\xi}$.
  Consequently, almost all energy levels of the ring Hamiltonian are degenerate.
  Only the ground state and -- for even $L$ -- the highest energy level are non-degenerate.

  The Schr\"odinger group tells us about the symmetry properties of the system {\em as a whole},
  but it is not the relevant object for the first detection problem.
  Consider the first detection of the localized state at the origin $\ket{\RDet} = \ket{0}$ on the ring.
  Any translation $\hat{T}_\xi$ will move the detection state, as will most reflections.
  The only symmetry transformations that leave $\ket{0}$ invariant are the identity $\Id$ and 
  the reflection around the origin $\hat{R}_0$.
  The group of all symmetry transformations that respect the Hamiltonian $\Ham$ {\em and} the detection 
  state is the stabilizer subgroup $\Stab \subseteq \mathcal{A}$:
  \begin{equation}
    \Stab = \{ \hat{S} \in \mathcal{A} \, | \, \hat{S} \ket{\RDet} = \ket{\RDet} \}
  \label{eq:DefStab}
  \end{equation}
  In our example, we have $\mathcal{S}_{\dyad{0}} = \{ \Id , \hat{R}_0 \}$, see the right-hand-side of Fig.~\ref{fig:SymmetryTransforms}.
  For each $\hat{S} \in \Stab$, we have $\hat{S}^\dagger \hat{S} = \Id$ (unitarity), $\comm*{\hat{S}}{\Ham} = 0$ 
  (system symmetry) and the stabilizing property $\hat{S} \ket{\RDet} = \ket{\RDet}$.
  The last equation also implies that $\hat{S}$ commutes with $\Detect$ and that $\bra{\RDet}\hat{S} = \bra{\RDet}$.

  The stabilizer formalizes the concept of physical equivalent states:
  {\em Two initial states $\ket{\psi}$ and $\ket{\psi'}$ are physically equivalent
  when there exists a symmetry operation $\hat{S} \in \Stab$ from the 
  stabilizer subgroup, such that}
  \begin{equation}
    \ket{\psi'} = \hat{S} \ket{\psi}
    .
  \label{eq:DefPhysEqSymm}
  \end{equation}
  This is equivalent to the requirement \eqref{eq:DefPhysEq} that $\ket{\psi}$ and $\ket{\psi'}$ give the same transition amplitudes
  to the detection state.
  Because if $\ket{\psi}$ and $\ket{\psi'} = \hat{S}\ket{\psi}$ are physically equivalent, we find:
  \begin{align}
    \mel*{\RDet}{\TEO(t)}{\psi'}
    = & \nonumber
    \mel*{\RDet}{\TEO(t)\hat{S}}{\psi}
    \\ = &
    \mel*{\RDet}{\hat{S}\TEO(t)}{\psi}
    =
    \mel*{\RDet}{\TEO(t)}{\psi}
    .
  \label{eq:}
  \end{align}
  The same trick can be used to pass $\hat{S}$ through all operators in the definition \eqref{eq:DefFDA},
  because $\comm*{\hat{S}}{\Detect} = 0$
  \begin{align}
    \FDA_n(\psi')
    = & \nonumber
    \mel*{\RDet}{\TEO(\tau) [(\Id - \Detect)\TEO(\tau)]^{n-1}\hat{S}}{\psi}
    \\ = & \nonumber
    \mel*{\RDet}{\hat{S} \TEO(\tau) [(\Id - \Detect)\TEO(\tau)]^{n-1}}{\psi}
    \\ = & 
    \mel*{\RDet}{\TEO(\tau) [(\Id - \Detect)\TEO(\tau)]^{n-1}}{\psi}
    = \FDA_n(\psi)
    .
  \end{align}
  For each localized initial state $\ket{\RIn}$ on the ring, we find that 
  $\ket{-\RIn} = \ket{L-\RIn} = \hat{R}_0 \ket{\RIn}$ is physically equivalent to it.
  However, the definitions made here are far more general.

  We denote the linear space of all physically equivalent initial states with $\mathcal{E}_{\ket{\PsiIn}} = \sSpan{\hat{S}\ket{\PsiIn} \, | \, \hat{S} \in \Stab }$.
  The {\em number of physically equivalent states} $\nu(\PsiIn)$ to some initial state can now 
  formally be defined as the dimension of the vector space that is generated by the stabilizer acting on 
  the initial state, i.e. the dimension of $\mathcal{E}_{\ket{\PsiIn}}$:\footnote{
    The set $\Stab \ket{\PsiIn}$ is called the ``orbit'' of $\ket{\PsiIn}$ under the 
    action of the stabilizer subgroup.
    Therefore the technical name of $\nu$ is the dimension of the $\ket{\RDet}$-stabilizer orbit of $\ket{\PsiIn}$.
    Furthermore, two initial states are physically equivalent, if and only if they share the same orbit.
  }
  \begin{equation}
    \nu(\PsiIn) 
    := 
    \mathrm{dim} \qty[\Span{\hat{S} \ket{\PsiIn} \, | \, \hat{S} \in \Stab }]
    .
  \label{eq:DefNuSymm}
  \end{equation}
  In our example, this subspace is $\mathcal{E}_{\ket{\RIn}} = \sSpan{ \Id \ket{\RIn} , \hat{R}_0 \ket{\RIn} } 
  = \sSpan{ \ket{\RIn} , \ket{-\RIn} }$, so that $\nu(\RIn) = 2$.
  An exception appears for rings of even lengths with the initial state $\ket{\RIn} = \ket{L/2}$.
  In this case we find $\hat{R}_0 \ket{L/2} = \ket{L/2}$, thus $\mathcal{E}_{\ket{L/2}} = \sSpan{ \ket{L/2}, \ket{L/2} }$,
  and therefore $\nu(L/2) = 1$.
  The definition \eqref{eq:DefNuSymm} is general in that it explicitly admits non-localized initial states.

  Having identified the stabilizer $\Stab$, it is clear what states are ``symmetric'': 
  those that are invariant under the stabilizer action.
  A projector onto all symmetric states is constructed via:
  \begin{equation}
    \hat{P}_{\mathcal{S}}
    :=
    \frac{1}{\abs{\Stab}} \Sum{\hat{S}\in\Stab}{} \hat{S}
    ,
  \label{eq:DefSymmProj}
  \end{equation}
  where $\abs{\Stab}$ denotes the cardinality of the stabilizer subgroup.
  Since the subgroup $\Stab$ is closed, the multiplication with any $\hat{S}'\in\Stab$ only reorders the terms in 
  the sum and does not change the result.
  Hence application of $\hat{P}_{\mathcal{S}}$ yields a symmetric state.
  For the same reason it is a projector, i.e. $\hat{P}_{\mathcal{S}}^2 = \hat{P}_{\mathcal{S}}$.
  Also, we have $\hat{P}_{\cal S} \ket{\RDet} = \ket{\RDet}$.

  The symmetry projector is used to define the AUS of an arbitrary initial state $\ket{\PsiIn}$ as its symmetric part.
  Consider an arbitrary initial state $\ket{\PsiIn}$ with $\nu$ physically equivalent sites $\ket{\psi_j} = \hat{S}_j \ket{\PsiIn}$, 
  where we can set $\ket{\psi_0} = \ket{\PsiIn}$ and $\hat{S}_0 = \Id$.
  The projector $\hat{P}_{\cal S}$ appears naturally when considering the sum $\sSum{j=0}{\nu-1} \hat{S}_j \ket{\PsiIn}$.
  The generalization of Eq.~\eqref{eq:DefAUS} therefore reads:
  \begin{equation}
    \ket{\AUS(\PsiIn)}
    :=
    \frac{
      \hat{P}_{\cal S} \ket{\PsiIn}
    }{
      \sqrt{\smash[b]{\ev*{\hat{P}_{\cal S}}{\PsiIn}}}
    }
    .
  \label{eq:DefAUSSymm}
  \end{equation}
  Since $\hat{P}_{\cal S}$ is a projector, we have $0 \le \ev*{\hat{P}_{\cal S}}{\PsiIn} \le 1$.
  Whenever $\ev*{\hat{S}}{\PsiIn} = 0$ for all $\hat{S}\ne\Id$, we have $\ev*{\hat{P}_{\cal S}}{\PsiIn} = 1/\nu$ and Eq.~\eqref{eq:DefAUS} is restored.
  [Note that each $\ket{\psi_j}$ appears $\abs{\Stab}/\nu$ times in the sum $\hat{P}_{\cal S}\ket{\PsiIn}$.]
  This is particularly the case for localized initial states $\ket{\PsiIn} = \ket{\RIn}$.
  A counter-example is the cross-structure of Fig.~\ref{fig3F} with the initial state $\ket{\PsiIn} = (\ket{1} + e^{i\alpha}\ket{2})/\sqrt{2}$.
  Here, $\hat{P}_{\cal S} \ket{\PsiIn} = (1 + e^{i\alpha}) \ket{\AUS}/\sqrt{2\nu}$ so that $\ev*{\hat{P}_{\cal S}}{\PsiIn} = (1 + \cos\alpha)/\nu$, 
  where $\nu=4$ and $\ket{\AUS}$ is the AUS of Fig.~\ref{fig3F}.
  In parallel to Eq.~\eqref{eq:AUSAndNu}, we have:
  \begin{equation}
    \TDP(\PsiIn)
    =
    \ev*{\hat{P}_{\cal S}}{\PsiIn}
    \TDP(\AUS(\PsiIn))
    .
  \label{eq:AUSAndSymm}
  \end{equation}

  Let us now make the connection to [I] and Eq.~\eqref{eq:Kessler}.
  The AUS is symmetric under the stabilizer action by construction.
  As a consequence, it must be a superposition of eigenstates that are also symmetric under the 
  stabilizer action.
  Which are the symmetric eigenstates?
  The answer has two parts:
  Firstly, all bright eigenstates $\ket{\beta_l}$ of Eq.~\eqref{eq:DefBrightEigen} are symmetric.
  Since $\hat{S} \in \Stab$ necessarily commutes not only with $\Ham$, but also with all
  of its eigenspace projectors $\hat{P}_l$, we have:
  \begin{equation}
    \hat{S}\ket{\beta_l}
    =
    N_l \hat{S} \hat{P}_l \ket{\RDet}
    =
    N_l \hat{P}_l \hat{S} \ket{\RDet}
    =
    N_l \hat{P}_l\ket{\RDet}
    =
    \ket{\beta_l}
    ,
  \label{eq:}
  \end{equation}
  where $N_l = [\ev*{\hat{P}_l}{\RDet}]^{-1/2}$.
  Each bright eigenstate is symmetric under the stabilizer action.
  However, there may also be symmetric dark states $\hat{P}_{\mathcal{S}}\ket*{\delta^{\mathcal{S}}_{l,m}} = \ket*{\delta^{\mathcal{S}}_{l,m}}$!
  In particular, every completely dark quasienergy level (i.e. any level with $\hat{P}_l\ket{\PsiDet} = 0$, 
  see [I]) will contain a symmetric dark state.

  According to Eq.~\eqref{eq:DefAUSSymm}, the AUS must be a superposition of bright eigenstates and of symmetric dark eigenstates:
  \begin{equation}
    \ket*{\AUS(\PsiIn)} 
    = 
    \sideset{}{'}\sum_{l} \ket{\beta_l}\ip{\beta_l}{\PsiIn}
    +
    \sideset{}{''}\sum_{l,m} \ket*{\delta^{\mathcal{S}}_{l,m}} \ip*{\delta^{\mathcal{S}}_{l,m}}{\PsiIn}
    .
  \label{eq:}
  \end{equation}
  Here, the first sum runs over all bright eigenstates, and the second sum runs over all symmetric stationary dark states.
  If there are no symmetric dark states, then $\ket{\AUS(\PsiIn)}$ must be bright.
  By the arguments laid out in [I], the AUS is then detected with probability one,
  the upper bound saturates and 
  \begin{equation}
    \TDP(\PsiIn)
    =
    \ev{\hat{P}_{\cal S}}{\PsiIn}
    ,
  \label{eq:}
  \end{equation}
  from Eq.~\eqref{eq:AUSAndSymm}.
  Depending on $\ket{\PsiIn}$ this may or may not be equal to $1/\nu(\PsiIn)$.

  Throughout this section we still assumed that resonant detection periods are avoided and that the detection state is localized.
  The general case is conceptually the same and discussed in appendix~\ref{app:Peculiarities}.
  In order to drop the first assumption, we need to amend Eq.~\eqref{eq:DefSchroedingerGroup} and require that the system's symmetries
  commute with $\TEO(\tau)$ instead with the Hamiltonian.
  $\mathcal{A}$ will not be evident from the graph structure alone anymore.
  When general detection states are considered, stabilizer symmetries may introduce a phase factor to the detection state.
  These phase factors will appear in all equations regarding physical equivalent states.
  They appear because the detection state may belong to a non-trivial representation of the stabilizer subgroup.
  As mentioned, all details and amendments are discussed in appendix~\ref{app:Peculiarities}.

  We have thus generalized our notion of physical equivalence and found the condition for the saturation of the upper bound:
  the lack of symmetric dark states.
  In the following section, we propose a dimensional reduction based on this insight.

\section{The symmetrized Hamiltonian}
\label{sec:SymHam}
  \begin{figure}
    \includegraphics[width=0.9\columnwidth]{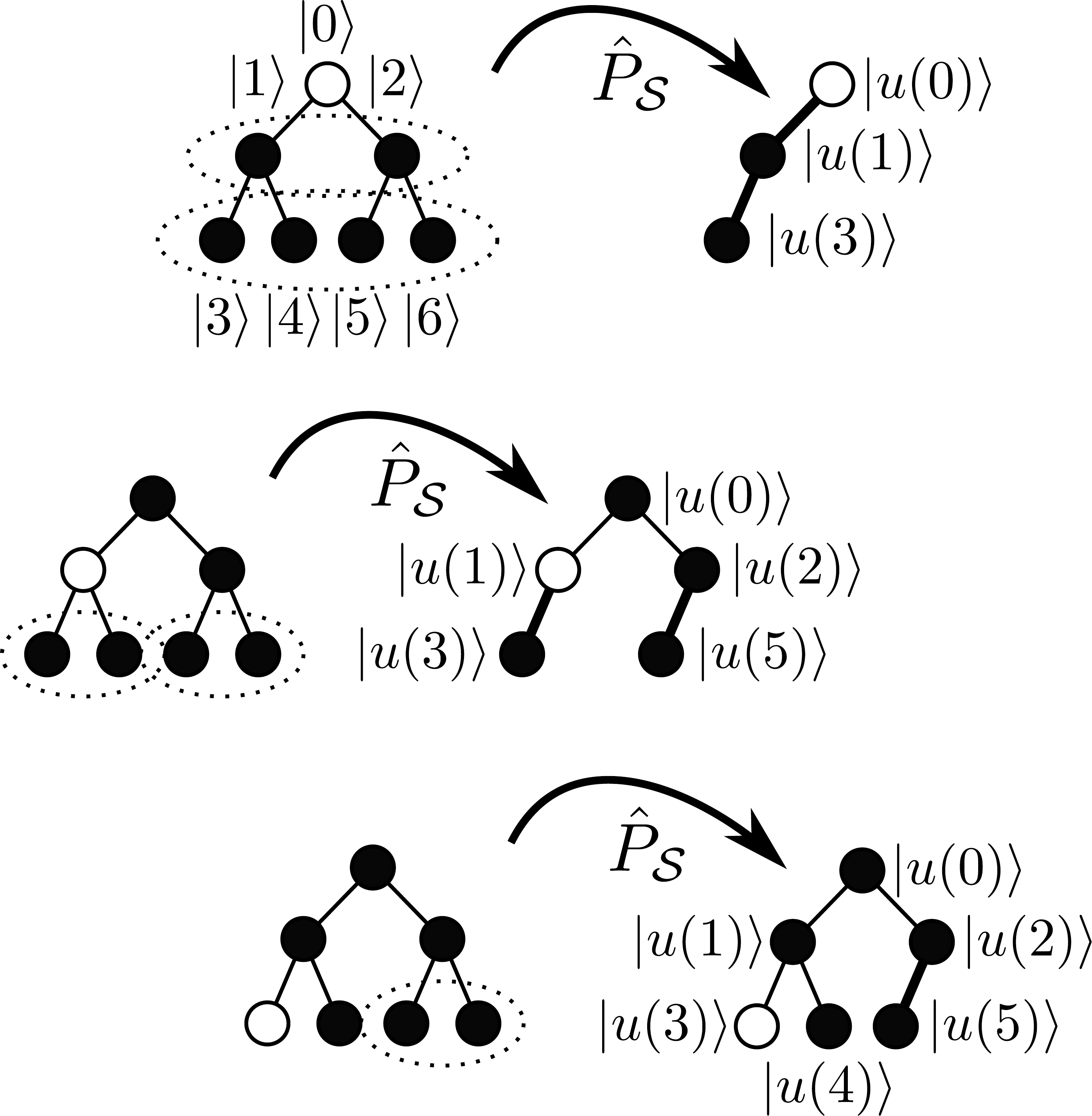}
    \caption{
      The symmetrized system for a small tree graph.
      Detection takes place on the site with the open circle.
      After symmetrization, we obtain another (weighted) graph system.
      All nodes which are physically equivalent (dotted groups) with respect to the detection site 
      are reduced to one node in the symmetrized system.
      Grouping multiple nodes may lead to different link strengths (thicker lines).
      (Here each thick line corresponds to a relative strength of $\sqrt{2}$.)
      The actual form of the symmetrized system depends on the location of the detector.
      Top: Detection takes place in the root.
      Middle: Detection in the middle.
      Bottom: Detection in the leaves.
      \label{fig:SymTree}
    }
  \end{figure}
  The important conclusion of sec.~\ref{sec:Stabilizer} is that the bright part of the Hilbert space $\Hilbert_B \subseteq \Hilbert_{\cal S} := \hat{P}_{\cal S} \Hilbert$ 
  is completely contained in the symmetric part $\Hilbert_{\mathcal{S}}$ of the Hilbert space.
  This allows for a reduction of dimensionality of the Hamiltonian.
  Naively, it is necessary to diagonalize the complete Hamiltonian $\Ham$ to find all eigenstates,
  which yield $\TDP$, see Eq.~\eqref{eq:Kessler} and [I].
  When the detection state is known and fixed, one may instead only diagonalize the ``symmetrized'' Hamiltonian $\Ham_{\cal S}$.
  This is obtained by replacing each localized state $\ket{r}$ with its AUS $\ket{\AUS(r)}$.
  So, when the position representation of the Hamiltonian is $\Ham = \sSum{x,y}{} h_{x,y} \dyad{x}{y}$,
  then the symmetrized Hamiltonian will be:
  \begin{equation}
    \Ham_{\cal S}
    :=
    \hat{P}_{\mathcal{S}} \Ham \hat{P}_{\mathcal{S}}
    =
    \Sum{\AUS(x), \AUS(y)}{}
    h_{\AUS(x),\AUS(y)}^{\cal S}
    \dyad{\AUS(x)}{\AUS(y)}
    .
  \label{eq:DefSymHam}
  \end{equation}
  Here the sum runs over all different equivalency classes, i.e. over all different AUSs.
  To obtain the new entries $h_{\AUS(x),\AUS(y)}^{\cal S}$, we make use of the definition \eqref{eq:DefAUSSymm} 
  and $\ev*{\hat{P}_{\cal S}}{x} = 1/\nu(x)$ for localized states, and sum over all physically equivalent sites $x'$ 
  appearing in $\AUS(x)$ (or $\AUS(y)$).
  \begin{equation}
    h_{\AUS(x),\AUS(y)}^{\cal S}
    :=
    \Sum{x' \sim \AUS(x)}{}
    \Sum{y' \sim \AUS(y)}{}
    \frac{h_{x',y'}}{\sqrt{ \nu(x') \nu(y') }}
    .
  \label{eq:DefSymmRates}
  \end{equation}
  This procedure does not necessarily lead to equal entries in the symmetrized Hamiltonian.
  That means, when $\Ham$ describes a quantum walk on a graph, then $\Ham_{\cal S}$ may describe a quantum walk on a {\em weighted} graph,
  where each link may have an individual strength.
  The new link strength depends on the number of physical equivalent partners of $x$ and $y$, as well as on the number of links between those two groups.
  The so-obtained graph is the quotient graph \cite{Krovi2007-0}.
  As an example, we consider the binary tree of [I], see Fig.~\ref{fig:SymTree}.
  Each group of physically equivalent states in the original graph is replaced by one node in the symmetrized graph which does not have equal link strengths everywhere.

  The benefit of this procedure is that there are no asymmetric states in $\Ham_{\cal S}$.
  The dimension of $\Ham_{\cal S}$ may therefore be significantly smaller than the one of the original Hamiltonian $\Ham$.
  Furthermore, the only dark states that can be found in the symmetrized Hamiltonian are the symmetric dark states.
  Consider the diagonal form of the symmetrized Hamiltonian:
  \begin{equation}
    \Ham_{\cal S}
    =
    \Sum{l}{} \Sum{m=1}{g_l^{\mathcal{S}}} E_l \dyad*{E_{l,m}^{\mathcal{S}}}
    .
  \label{eq:SymHamDiag}
  \end{equation}
  This consists exclusively out of symmetric eigenstates $\ket*{E_{l,m}^{\cal S}}$, which are superpositions of 
  different AUSs $\ket{\AUS(r)}$ instead of individual position eigenstates $\ket{r}$.
  Obviously, the symmetrization procedure changes the degeneracy of each energy level.
  $E_l$ may have $g_l$ different eigenstates, but it only has $1 \le g_l^{\mathcal{S}} \le g_l$ symmetric eigenstates.
  Any energy level that is still degenerate, even after the symmetrization procedure, 
  must possess a symmetric dark state, which ruins the equality of the upper bound.
  (So does every energy level that has no overlap with the detection state at all.)
  The bright energy eigenstates $\ket{\beta_l}$ are composed out of symmetric energy eigenstates only.
  Therefore, applying the whole machinery of [I] to the Hamiltonian \eqref{eq:SymHamDiag} will yield the same result for 
  the total detection probability:
  \begin{equation}
    \TDP(\PsiIn)
    =
    \sideset{}{'} \sum_l
    \frac{
      \abs*{ \sSum{m=1}{g_l^{\mathcal{S}}}
        \ip*{\PsiDet}{E_{l,m}^{\mathcal{S}}} 
        \ip*{E_{l,m}^{\mathcal{S}}}{\PsiIn} 
      }^2
    }{
      \sSum{m=1}{g_l^{\cal S}}
      \ip*{\PsiDet}{E_{l,m}^{\mathcal{S}}} 
      \ip*{E_{l,m}^{\mathcal{S}}}{\PsiDet} 
    }
    .
  \label{eq:SymKessler}
  \end{equation}
  As in [I], the sum excludes all completely dark energy levels, such that the denominator can never vanish.
  The advantage of this equation is that $\Ham_{\mathcal{S}}$ may have a much smaller dimension than the original Hamiltonian.
  Therefore, the eigenstates of $\Ham_{\mathcal{S}}$ are easier to obtain.
  Its disadvantage is that it is specific to the exact detection state $\ket{\PsiDet}$ through the stabilizer subgroup.
  Consequently, if one wants to obtain $\TDP$ for a different detection state, one must start with the computation of the new $\Ham_{\mathcal{S}}$.

  Let us demonstrate this program in the tree example of before.

  \subsection{The tree graph}
    The original Hamiltonian of the tree graph in Fig.~\ref{fig:PD} is given by:
    \begin{equation}
      \Ham
      =
      - \gamma \mqty(
        0 & 1 & 1 & 0 & 0 & 0 & 0 \\
        1 & 0 & 0 & 1 & 1 & 0 & 0 \\
        1 & 0 & 0 & 0 & 0 & 1 & 1 \\
        0 & 1 & 0 & 0 & 0 & 0 & 0 \\
        0 & 1 & 0 & 0 & 0 & 0 & 0 \\
        0 & 0 & 1 & 0 & 0 & 0 & 0 \\
        0 & 0 & 1 & 0 & 0 & 0 & 0 
      )
      ,
    \label{eq:HamTree}
    \end{equation}
    where $\gamma$ is an energy constant.
    The graph has seven nodes:
    the root $\ket{0}$, two nodes in the middle $\ket{1}$ and $\ket{2}$, and the leaves $\ket{3}, \ket{4}, \ket{5}$, and $\ket{6}$.
    See Fig.~\ref{fig:SymTree} for the notation.
    Clearly, the total symmetry group of that tree is generated by three operations:
    $\hat{A}_1 := \dyad{3}{4} + \dyad{4}{3}$ that switches the left pair of leaves, $\hat{A}_2 := \dyad{5}{6} + \dyad{6}{5}$ 
    that switches the right pair and $\hat{A}_0 := \dyad{1}{2} + \dyad{3}{5} + \dyad{4}{6} + \mathrm{h.c.}$, that switches the 
    left and right sub-trees.
    As $\hat{A}_0$ does not commute with the other two generators, we find $\abs{\mathcal{A}} = 8$.

    The stabilizer $\Stab$ and thus $\Ham_{\cal S}$ depend on the choice of the detection node.
    Still one finds that $\Ham_{\cal S}$ can be represented by a weighted graph.
    This is represented in Fig.~\ref{fig:SymTree}.
    We discuss all different choices of localized detection states in the following.
    For reasons of space, we write $\ket{\AUS_r} := \ket{\AUS(r)}$ for the remainder of this section.

    \paragraph{Detection in the leaves.}
      We set $\ket{\RDet} = \ket{3}$, see Fig.~\ref{fig:SymTree}.
      The stabilizer subgroup contains only $\hat{A}_2$ and the identity, i.e. $\Stab = \{ \Id , \hat{A}_2 \}$.
      Hence $\ket{5}$ and $\ket{6}$ are physically equivalent with $\nu(5) = \nu(6) = 2$ and all other states are unique.
      The right pair of leaves becomes a pair of physically equivalent nodes, that is mapped to
      one AUS $\ket{u_5} = (\ket{5} + \ket{6})/\sqrt{2}$.
      Remember that the third node is the detection node: $\ket{u_3} = \ket{3} = \ket{\RDet}$.
      Each other node is unique and represents its own AUS $\ket{u_j} = \ket{j}$, $j=0,\hdots,4$.
      The symmetrized Hamiltonian is written as a matrix for the AUS states using Eq.~\eqref{eq:DefSymmRates}.
      This matrix is given by:
      \begin{equation}
        \Ham_{\mathcal{S}} 
        =
        - \gamma
        \mqty(
          0 & 1 & 1 & 0 & 0 & 0 \\
          1 & 0 & 0 & 1 & 1 & 0 \\
          1 & 0 & 0 & 0 & 0 & \sqrt{2} \\
          0 & 1 & 0 & 0 & 0 & 0 \\
          0 & 1 & 0 & 0 & 0 & 0 \\
          0 & 0 & \sqrt{2} & 0 & 0 & 0 \\
        )
        ,
      \label{eq:}
      \end{equation}
      and corresponds to a weighted graph with one ``heavier'' link, see Fig.~\ref{fig:SymTree}.
      The symmetrized energy levels are $E_1 = -2\gamma$, $E_2 = - \sqrt{2}\gamma$, $E_3 = 0$, $E_4 = \sqrt{2}\gamma$, and $E_5 = 2\gamma$, of which $E_3$ is twice degenerate:
      \begin{equation}
        \left\{ \begin{aligned}
          \ket*{E_1^\mathcal{S}}
          = & 
          \frac{2 \ket{u_0} + 2 \ket{u_1} + 2 \ket{u_2} + \ket{u_3} + \ket{u_4} + \sqrt{2}\ket{u_5}}{4} 
          \\
          \ket*{E_2^\mathcal{S}}
          = & 
          \frac{2 \ket{u_1} - 2 \ket{u_2} + \sqrt{2} \ket{u_3} + \sqrt{2} \ket{u_4} - 2 \ket{u_5}}{4}
          \\
          \ket{E_{3,1}^\mathcal{S}}
          = &
          \frac{\sqrt{2}\ket{u_0}  - \sqrt{2} \ket{u_3}  - \ket{u_5}}{\sqrt{5}}
          \\
          \ket{E_{3,2}^\mathcal{S}}
          = &
          \frac{\ket{u_3} - \ket{u_4}}{\sqrt{2}}
          \\
          \ket{E_4^\mathcal{S}}
          = & 
          \frac{2 \ket{u_1} - 2 \ket{u_2} - \sqrt{2} \ket{u_3} - \sqrt{2} \ket{u_4} + 2 \ket{u_5}}{4}
          \\
          \ket*{E_5^\mathcal{S}}
          = & 
          \frac{2 \ket{u_0} - 2 \ket{u_1} - 2 \ket{u_2} + \ket{u_3} + \ket{u_4} + \sqrt{2}\ket{u_5}}{4} 
        \end{aligned} \right.
      \end{equation}
      As $E_3$ is twice degenerate, it yields a symmetric dark state.
      We plug the symmetric energy eigenstates into Eq.~\eqref{eq:SymKessler} to recover the result from [I]:
      \begin{equation}
        \TDP(\RIn)
        =
        \left\{ \begin{aligned}
          \tfrac{3}{5}\qc & \RIn = 0,4 \\
          1\qc & \RIn=1,2,3 \\
          \tfrac{2}{5}\qc & \RIn=5,6
        \end{aligned} \right.
        .
      \label{eq:}
      \end{equation}
      A little attention has to be paid for the right pair of leaves, as we find $\TDP(u_5) = 4/5$ but $\TDP(5) = \TDP(u_5)/2$,
      and similarly for $\ket{\RIn} = \ket{6}$.

    \paragraph{Detection in the middle.}
      We now pick $\ket{\RDet} = \ket{1} = \ket{u_1}$.
      The allowed symmetry transformations in the stabilizer are $\Stab = \{ \Id, \hat{A}_1, \hat{A}_2, \hat{A}_1\hat{A}_2 \}$.
      Both pairs of leaves get mapped to one AUS each: $\ket{u_3} = (\ket{3} + \ket{4})/\sqrt{2}$ and $\ket{u_5} := (\ket{5} + \ket{6})/\sqrt{2}$.
      The remaining nodes are AUSs themselves.
      The symmetrized Hamiltonian reads:
      \begin{equation}
        \Ham_{\cal S}
        =
        - \gamma \mqty(
          0 & 1 & 1 & 0 & 0 \\
          1 & 0 & 0 & \sqrt{2} & 0 \\
          1 & 0 & 0 & 0 & \sqrt{2} \\
          0 & \sqrt{2} & 0 & 0 & 0 \\
          0 & 0 & \sqrt{2} & 0 & 0 
        )
        .
      \label{eq:}
      \end{equation}
      The energy levels are the same as before, but none of them is degenerate as it was the case before:
      \begin{equation}
        \left\{ \begin{aligned}
          \ket*{E_1^\mathcal{S}}
          = & 
          \frac{2 \ket{u_0} + 2 \ket{u_1} + 2 \ket{u_2} + \sqrt{2} \ket{u_3} + \sqrt{2}\ket{u_5}}{4} 
          \\
          \ket*{E_2^\mathcal{S}}
          = & 
          \frac{\ket{u_1} - \ket{u_2} + \ket{u_3} - \ket{u_5}}{2}
          \\
          \ket{E_3^\mathcal{S}}
          = &
          \frac{\sqrt{2}\ket{u_0}  -\ket{u_3} - \ket{u_5}}{2}
          \\
          \ket{E_4^\mathcal{S}}
          = & 
          \frac{\ket{u_1} - \ket{u_2} - \ket{u_3} + \ket{u_5}}{2}
          \\
          \ket*{E_5^\mathcal{S}}
          = & 
          \frac{2 \ket{u_0} - 2 \ket{u_1} - 2 \ket{u_2} + \sqrt{2} \ket{u_3} + \sqrt{2}\ket{u_5}}{4} 
        \end{aligned} \right.
      \end{equation}
      Although there are no degenerate symmetric energy levels, $\ket*{E_3^{\cal S}}$ has no 
      overlap with $\ket{\RDet} = \ket{u_1}$ and is a dark state.
      Hence the upper bound is not an equality.
      Eq.~\eqref{eq:SymKessler} recovers the result from [I]:
      \begin{equation}
        \TDP(\RIn)
        =
        \left\{ \begin{aligned}
          \tfrac{1}{2} \qc & \RIn = 0 \\
          1 \qc & \RIn = 1,2 \\
          \tfrac{3}{8}\qc & \text{otherwise}
        \end{aligned} \right.
        .
      \label{eq:}
      \end{equation}

    \paragraph{Detection on the root.}
      Finally, we choose $\ket{\RDet} = \ket{0}$.
      The stabilizer with respect to this detection state is the full symmetry group of the tree.
      There are three uniform states, that group all nodes of each generation:
      $\ket{u_0} = \ket{0} = \ket{\RDet}$, $\ket{u_1} = (\ket{1} + \ket{2})/\sqrt{2}$, and $\ket{u_3} = (\ket{3} + \ket{4} + \ket{5} + \ket{6})/2$.
      The symmetrized graph becomes a line, see Fig.~\ref{fig:SymTree}, with the associated Hamiltonian:
      \begin{equation}
        \Ham_{\cal S} 
        =
        - \gamma \mqty(
          0 & \sqrt{2} & 0 \\
          \sqrt{2} & 0 & \sqrt{2} \\
          0 & \sqrt{2} & 0
        )
        .
      \label{eq:}
      \end{equation}
      This Hamiltonian is three-dimensional, in contrast to the original seven-dimensional one of Eq.~\eqref{eq:HamTree}.
      It has three non-degenerate energy levels, which all overlap with the detection state.
      Hence, there are no symmetric dark states, and the upper bound is exact.
      We find:
      \begin{equation}
        \TDP(\RIn)
        =
        \left\{ \begin{aligned}
          1 \qc & \RIn = 0 \\
          \tfrac{1}{2} \qc & \RIn = 1,2 \\
          \tfrac{1}{4} \qc & \RIn = 3,4,5,6
        \end{aligned} \right.
        ,
      \label{eq:}
      \end{equation}
      just as in [I].

\section{Summary and discussion}
\label{sec:Summary}
  In this series, we are concerned with the overall probability $\TDP$ to find a quantum particle in 
  some target state $\ket{\PsiDet}$ under stroboscopic measurement with frequency $1/\tau$, when it was initially in state $\ket{\PsiIn}$.
  Here, we investigated the influence of the system's symmetries on $\TDP$.
  Whenever one has found $\nu$ physically equivalent initial states, one will find that
  the total detection probability is bounded by $\TDP \le 1/\nu$.
  Two states are physically equivalent when they yield the same transition amplitudes to the detection
  state for all times.
  Any pair of equivalent states can be seen to yield a dark state from their negative superposition.
  The AUS defined by Eq.~\eqref{eq:DefAUS} is the positive superposition of all equivalent states, contains all bright components of the
  original initial state, and gives $\TDP$ via Eq.~\eqref{eq:AUSAndNu}.

  The stabilizer subgroup $\Stab$ of the system's symmetry group respects the system dynamics and 
  the detection process.
  It formalizes the notion of physical equivalence and yields the more 
  general definitions (\ref{eq:DefNuSymm},\ref{eq:DefAUSSymm}) of $\nu$ and the AUS.
  $\nu(\PsiIn)$ of Eq.~\eqref{eq:DefNuSymm} is the dimension of the space of all states 
  physically equivalent to $\ket{\PsiIn}$.
  The AUS is the stabilizer-symmetric component of the initial state.
  Also all bright energy eigenstates are stabilizer-symmetric.
  When the system has no additional symmetric dark states, then the AUS consists only of bright states
  and the upper bound Eq.~\eqref{eq:UpperBound} saturates.
  Diagonalizing the symmetrized Hamiltonian is a possible method to reduce the 
  dimensionality of the original problem and yields the exact formula \eqref{eq:SymKessler} for $\TDP$.

  It is tempting to characterize the Schr\"odinger group and the stabilizer subgroup
  by its irreducible representations, similar to the discussion in Ref.~\cite{Krovi2007-0}.
  Knowledge about the irreducible representations of $\Stab$ can then give sufficient conditions for the 
  existence of dark states and for possible deficits in $\TDP$.
  We found this programme not practical because it does not take accidental degeneracies into account.
  However, ``accidental'' degeneracies appear quite frequently, for example in the binary tree of Eq.~\eqref{eq:HamTree}.
  The real culprit responsible for dark states is degeneracy and not symmetry.
  Degeneracies are not completely accessible from symmetry considerations alone.
  Still, a connection to the one-dimensional (trivial or non-trivial) irreducible representations of $\Stab$ can be made 
  when the stabilizer action on the detection space is considered, see appendix~\ref{app:Peculiarities}.

  The upper bound is our main finding in this article and we would like to stress 
  its importance, which is not only due to how easy it is obtained.
  The inequality~\eqref{eq:UpperBound} is valid even when one can not identify all equivalent states.
  If one can only find $\nu_\text{app}$ apparent equivalent states, although there are $\nu_\text{true} > \nu_\text{app}$ 
  in the system, $\TDP \le 1/\nu_\text{app}$ is still true.
  When the detection period $\tau$ hits a resonant value, Eq.~\eqref{eq:DefResonantTau}, and $\TDP$ drops, see [I], 
  then $\nu_\text{true}$ increases, due to additional, dynamical symmetries in the evolution operator
  that are not accessible from an inspection of the Hamiltonian alone; yet $\TDP \le 1/\nu_\text{app}$.
  We expect the upper bound to be valid as well when there are irregularities in the periods between detection attempts.
  We base this belief on the fact that neither $\TDP$ nor $\nu$ depend on $\tau$ apart from resonant values
  which are of measure zero.
  Finally, the upper bound is still valid in infinite-dimensional systems, when the theory of [I] 
  breaks down and $\TDP$ actually does become a non-trivial function of $\tau$, see Refs.~\cite{Friedman2017-1, Thiel2018-0}.
  The only ingredient for the upper bound are the dark states that can be constructed from physically equivalent partners;
  and these are present as well in infinite-dimensional systems.

  \acknowledgements
    The support of Israel Science Foundation's grant 1898/17 is acknowledged.
    FT is supported by DFG (Germany) under grant TH 2192/1-1.

\appendix
\section{Details on the stabilizer subgroup}
\label{app:Peculiarities}
  Throughout the main text, we made two assumptions.
  First we assumed that $\tau$ is not equal to one of the resonant values of Eq.~\eqref{eq:DefResonantTau}.
  Secondly, we assumed the detection state to be localized.
  Both assumptions will now be dropped, and the necessary amendments will be explained.
  As an example we will again consider the ring with $L$ sites and the Hamiltonian \eqref{eq:TBHam}.

  \subsection{Possibly resonant detection periods}
    Resonant detection periods are defined by Eq.~\eqref{eq:DefResonantTau}.
    At these particular values $\tau = \tau_c$, two (or more) energy levels, say $E_l$ and $E_{l'}$,
    become dynamically equivalent.
    This means that, although they belong to different eigenvalues in the Hamiltonian, they belong to the 
    {\em same} eigenvalue in the evolution operator, namely $e^{-i\lambda}$, where $\lambda = \tau E_l = \tau E_{l'} \mod 2\pi$.
    This leads to additional degeneracies in $\TEO(\tau_c)$ that are not present in $\Ham$.

    The additional degeneracy corresponds to additional symmetries that are present in $\TEO(\tau_c)$, but not in $\Ham$.
    The system's symmetry group can thus be defined in a robust way, by requiring the operations to commute with $\TEO(\tau)$ 
    instead of with the Hamiltonian.
    This means, we replace Eq.~\eqref{eq:DefSchroedingerGroup} with the more general version:
    \begin{equation}
      \mathcal{A}
      :=
      \{ \hat{A} \text{ is unitary} \, | \, \comm*{\hat{A}}{\TEO(\tau)} = 0 \}
      .
    \label{eq:AppDefSchroedingerGroup}
    \end{equation}
    Obviously $\mathcal{A}$ changes when $\tau$ hits a resonant value, but otherwise it is the same as Eq.~\eqref{eq:DefSchroedingerGroup}.
    Consequently, the stabilizer subgroup may also be larger for resonant $\tau$ than it is for non-resonant values.
    This means, there are more dark states in the system, and more physically equivalent states.

    Our first definition of physically equivalent states \eqref{eq:DefPhysEq} can safely be relaxed.
    Instead of requiring equal transition amplitudes to the detection state for {\em all} times, it is sufficient
    to have equal transition amplitudes at the {\em right} times.
    Replacing Eq.~\eqref{eq:DefPhysEq}, we call two initial states $\ket{\PsiIn}$ and $\ket{\PsiIn'}$ physically equivalent when
    \begin{equation}
      \mel*{\PsiDet}{\TEO(n\tau)}{\PsiIn}
      =
      \mel*{\PsiDet}{\TEO(n\tau)}{\PsiIn'}
      \qc \text{for any } n
    \label{eq:AppDefPhysEq1}
    \end{equation}
    This condition is specific to the stroboscopic  protocol, while the condition of the main text
    will also hold for any variation in the sequence of measurement times.

    As an example consider the ring with $L=6$ sites, which has the energy levels $-2\gamma$, $-\gamma$, $\gamma$, and $2\gamma$.
    The spectrum is commensurable and the system admits a complete revival at the time $\tau_c = (\pi/2) (\hbar/\gamma)$.
    At this detection period all energy levels collapse to one quasi-energy level $\lambda=\pi$.
    The evolution operator for this detection frequency is the identity: $\TEO(\tau_c) = \Id$.
    The dynamics are trivial, we find $\FDA_n(\PsiIn) = \ip*{\PsiDet}{\PsiIn} \delta_{n,1}$ from Eq.~\eqref{eq:DefFDA}, 
    and $\TDP(\PsiIn) = \abs{\ip*{\PsiDet}{\PsiIn}}^2$.

    At the resonance, it becomes difficult to infer $\mathcal{A}$ from the {\em Hamiltonian's} graph structure.
    $\nu$ will increase and $\TDP(\PsiIn)$ will drop.
    As discussed, our original upper bound Eq.~\eqref{eq:UpperBound} will still be correct.
    Furthermore it was demonstrated in Refs.~\cite{Gruenbaum2013-0, Friedman2017-0, Friedman2017-1, Yin2019-0}, 
    that the variance of the first detection time diverges when $\tau$ approaches a critical value.
    This blow-up of the first detection statistics will make the experimental detection of the resonances very easy, but 
    it will be difficult to investigate the statistics directly on resonance.
    These resonances are singular in nature and we expect them to disappear when any kind of non-commensurable
    variability is introduced into the times between detection attempts.

  \subsection{Stabilizer symmetries for general detection states}
    \subsubsection{General definitions of the stabilizer, physical equivalence and the symmetry projector}
      When $\ket{\PsiDet}$ is not a localized state on a graph, 
      but a general state in a general Hilbert space, the discussion of sec.~\ref{sec:Stabilizer}
      has to be slightly amended.
      The reason is that stabilizer symmetries are allowed to change the {\em phase} of the detection state.
      This way, they change the detection state, but they do stabilize the {\em detection space}.
      Therefore an admissible stabilizer symmetry is represented by a unitary operator $\hat{S}$ 
      that acts on the detection state like $\hat{S}\ket{\PsiDet} = e^{i\lambda(\hat{S})}\ket{\PsiDet}$.
      The detection bra obtains the same phase factor: $\bra{\PsiDet}\hat{S} = [\hat{S}^\dagger\ket{\PsiDet}]^\dagger = e^{i\lambda(\hat{S})} \bra{\PsiDet}$.
      Although $\ket{\PsiDet}$ itself is changed, the detection operator $\Detect = \dyad{\PsiDet}$ is not,
      because $\hat{S}\Detect\hat{S}^\dagger = e^{i\lambda(\hat{S}) - i\lambda(\hat{S})} \Detect = \Detect$.
      Alternatively, we may write $\comm*{\Detect}{\hat{S}} = 0$.
      Eq.~\eqref{eq:DefStab} needs to be replaced with the more general version:
      \begin{equation}
        \Stab
        :=
        \{ \hat{S} \in \mathcal{A} \, | \, \comm*{\Detect}{\hat{S}} = 0 \}
        ,
      \label{eq:appDefStab}
      \end{equation}
      where $\mathcal{A}$ is given by Eq.~\eqref{eq:AppDefSchroedingerGroup} or like in the main text.
      The stabilizer consists of all symmetry operations that respect the temporal evolution and the detection projector.

      We still call two states $\ket{\psi}$ and $\ket{\psi'}$ physically equivalent, 
      if and only if there exists  $\hat{S} \in \Stab$, such that $\ket{\psi'} = \hat{S} \ket{\psi}$.
      However Eq.~\eqref{eq:DefPhysEq} [or Eq.~\eqref{eq:AppDefPhysEq1}] is no longer valid, as the transition amplitudes to the detection 
      state may have obtained a spurious phase factor.
      Physically equivalent states do not have identical transition amplitudes to the detection state,
      but rather identical transition {\em probabilities}.
      This is because any $\hat{S} \in \Stab$ commutes with $\TEO(\tau)$ and with $\Detect$ and is unitary.
      The transition probability in $n$ steps is given by $\abs*{\mel*{\PsiDet}{\TEO(n\tau)}{\psi'}}^2 = \norm*{\Detect\TEO(n\tau)\ket{\psi'}}^2$.
      Using the definition of the physical equivalent state $\ket{\psi'}$ and the commutation properties
      of $\hat{S}$, as well as its unitarity, we arrive at:
      \begin{align}
      \abs*{\mel*{\PsiDet}{\TEO(n\tau)}{\psi'}}^2
        = & \nonumber
        \norm*{\Detect\TEO(n\tau)\hat{S}\ket{\psi}}^2
        =
        \norm*{\hat{S}\Detect\TEO(n\tau)\ket{\psi}}^2
        \\ = &
        \norm*{\Detect\TEO(n\tau)\ket{\psi}}^2
        =
        \abs*{\mel*{\PsiDet}{\TEO(n\tau)}{\psi}}^2
        .
      \label{eq:}
      \end{align}
      The same trick is applied to the first detection probabilities $F_n(\psi') := \abs*{\FDA_n(\psi')}^2 = \norm*{\Detect \TEO(\tau) [ ( \Id - \Detect) \TEO(\tau)]^{n-1}\ket{\psi'}}^2$:
      \begin{align}
        \FDP_n(\psi')
        = & \nonumber
        \norm*{\Detect \TEO(\tau) [(\Id - \Detect)\TEO(\tau)]^{n-1}\hat{S}\ket{\psi}}^2
        \\ = & \nonumber
        \norm*{\hat{S}\Detect \TEO(\tau) [(\Id - \Detect)\TEO(\tau)]^{n-1}\ket{\psi}}^2
        \\ = & \nonumber
        \norm*{\Detect \TEO(\tau) [(\Id - \Detect)\TEO(\tau)]^{n-1}\ket{\psi}}^2
        = & 
        \FDP_n(\psi)
        .
      \end{align}
      Therefore physically equivalent states have identical detection probabilities.

      The number $\nu$ of physically equivalent states is still defined in the same way as before,
      i.e. Eq.~\eqref{eq:DefNuSymm} carries over.
      However, the symmetry projector has to be defined more carefully now.
      The different phase factors $e^{i\lambda(\hat{S})}$ need to be taken into account, so that the detection state is 
      invariant under the action of this projector, $\hat{P}_{\cal S} \ket{\PsiDet} = \ket{\PsiDet}$.
      This is achieved by replacing Eq.~\eqref{eq:DefSymmProj} with 
      \begin{equation}
        \hat{P}_{\cal S}
        :=
        \frac{1}{\abs{\Stab}} \Sum{\hat{S} \in \Stab}{} 
        e^{-i\lambda(\hat{S})} \hat{S}
        .
      \label{eq:appDefSymmProj}
      \end{equation}
      The definition \eqref{eq:DefAUSSymm} of the AUS remains untouched, if the projector from 
      Eq.~\eqref{eq:appDefSymmProj} is used.
      To see how $\TDP(\AUS(\PsiIn))$ and $\TDP(\PsiIn)$ are related, one notes that 
      $\FDA_n(\hat{S}\ket{\PsiIn}) = e^{i\lambda(\hat{S})}\FDA_n(\PsiIn)$.
      This gives 
      \begin{equation}
        \FDA_n(\AUS(\PsiIn))
        =
        \frac{\FDA_n(\PsiIn)}{\sqrt{ \ev*{\hat{P}_{\cal S}}{\PsiIn} }} 
        \underbrace{
          \frac{1}{\abs{\Stab}} 
          \Sum{\hat{S}\in\Stab}{} e^{-i\lambda(\hat{S}) + i \lambda(\hat{S})}
        }_{= 1}
        ,
      \label{eq:appDefAUS}
      \end{equation}
      which again leads to Eq.~\eqref{eq:AUSAndSymm}.
      For some initial states, we may again find that $\ev*{\hat{P}_{\cal S}}{\PsiIn} = 1/\nu(\PsiIn)$.

    \subsubsection{Example}
      Let us again discuss the example of the ring with an even number $L$ of sites.
      The energy eigenstates are given by free wave states:
      \begin{equation}
        \ket{E_k} 
        :=
        \frac{1}{\sqrt{L}} \Sum{x=1}{L}
        e^{i\frac{2\pi}{L} kx} \ket{x}
        ,
      \label{eq:}
      \end{equation}
      where $k \in \{ -(L/2 -1), \hdots, -1, 0, 1, \hdots, L/2 \}$.
      The two states $\ket{E_k}$ and $\ket{E_{-k}}$ belong to the same energy levels,
      except for $\ket{E_0}$ and $\ket{E_{L/2}}$ which are non-degenerate.

      One of these energy eigenstates is chosen as a detection state, $\ket{\PsiDet} = \ket{E_{k_\text{d}}}$.
      Since then $\Detect$ commutes with $\TEO(\tau)$, we find $\FDA_n(\PsiIn) = \delta_{n,1} \ip*{E_{k_\text{d}}}{\PsiIn}$ 
      in view of Eq.~\eqref{eq:DefFDA}.
      Therefore, the total detection probability is trivially given by:
      \begin{equation}
        \TDP(\PsiIn)
        =
        \abs*{\ip*{E_{k_\text{d}}}{\PsiIn}}^2
        \qc
        \TDP(\RIn) = \frac{1}{L}
        .
      \label{eq:}
      \end{equation}
      For a localized initial state $\ket{\PsiIn} = \ket{\RIn}$, the result suggests that $\nu(\RIn) = L$.

      Let us determine the stabilizer subgroup.
      The symmetry operations of $\mathcal{A}$ are translations and reflections, defined in Eq.~\eqref{eq:DefSymmOpRing}.
      Their action on the energy eigenstates is given by:
      \begin{align}
        \hat{T}_\xi \ket{E_k} = e^{-i\frac{2\pi}{L} k \xi} \ket{E_k}
        \qc &
        \hat{R}_r \ket{E_k} = e^{i\frac{2\pi}{L} 2 r k} \ket{E_{-k}}
        \\
        \hat{R}_r \ket{E_0} = \ket{E_0}
        \qc &  
        \hat{R}_r \ket{E_{L/2}} = \ket{E_{L/2}}
        .
      \label{eq:}
      \end{align}
      This shows that the translations are stabilizing symmetries for all eigenstates, 
      but the reflections only stabilize the states $\ket{E_0}$ and $\ket{E_{L/2}}$.

      Consider first the case when $k_\text{d}$ is neither zero nor $L/2$.
      Reflections do not stabilize the detection state and we have $\Stab = \{ \hat{T}_1, \hat{T}_2, \hdots, \hat{T}_{L-1}, \hat{T}_L = \Id \}$.
      With any localized initial state, we find that $\nu = L$.
      $\TDP(\RIn) = 1/\nu$ is indeed confirmed by Eq.~\eqref{eq:AUSAndSymm}, because
      \begin{align}
        \hat{P}_{\cal S}\ket{\RIn}
        = & \nonumber
        \frac{1}{L} \Sum{x=1}{L} 
        e^{i \frac{2\pi}{L} k_\text{d} x} \hat{T}_x \ket{\RIn}
        =
        \frac{e^{-i\frac{2\pi}{L}k_\text{d}\RIn}}{L}
        \Sum{x'=1}{L}
        e^{i \frac{2\pi}{L} k_\text{d} x'} \ket{x'}
        \\ = &
        \frac{e^{-i\frac{2\pi}{L}k_\text{d}\RIn}}{\sqrt{L}}
        \ket{E_{k_\text{d}}}
        ,
      \label{eq:}
      \end{align}
      resulting in $\ket{\AUS(\RIn)} = \ket{E_{k_\text{d}}} = \ket{\PsiDet}$ and $\ev*{\hat{P}_{\cal S}}{\RIn} = 1/L$.
      For any other non-localized initial condition, we can write $\ket{\PsiIn} = \sSum{x=1}{L} \psi_x \ket{x}$ and use above result to find $\TDP(\PsiIn) = \abs*{\ip*{\PsiIn}{E_{k_\text{d}}}}^2$, just as mentioned before.

      Consider now the case that $k_\text{d} = L/2$.
      ($k_\text{d} = 0$ yields no phase factors and gives the situation of the main text.)
      Then the whole symmetry group stabilizes the detection state.
      The symmetry projector reads:
      \begin{equation}
        \hat{P}_{\cal S}
        =
        \frac{1}{2L} \Sum{x=1}{L} [ \hat{T}_x + \hat{T}_x\hat{R}_0 ]
        .
      \label{eq:}
      \end{equation}
      (All symmetry elements have the form $\hat{T}_x$ or $\hat{R}_0\hat{T}_x$ for some $x$.)
      Using localized initial conditions, we see that 
      \begin{align}
        \hat{P}_{\cal S}\ket{\RIn}
        = & \nonumber 
        \frac{1}{2L} \Sum{x=1}{L} e^{i \frac{2\pi}{L}\frac{L}{2} x} 
        \qty[ \ket{x + \RIn} + \ket{x - \RIn} ]
        \\ = &
        \frac{(-1)^{\RIn}}{\sqrt{L}}
        \ket{E_{L/2}}
        ,
      \end{align}
      which will give the exact same results as before.
      Furthermore, the additional reflection symmetry does not yield any more physically equivalent states,
      so that $\nu = L$, just as for other values of $k_\text{d}$.

    \subsubsection{Character theory}
      The phase factor can be replaced by the following bulky expression:
      \begin{equation}
        e^{i\lambda(\hat{S})}
        =
        \ev{\hat{S}}{\PsiDet}
        =
        \Trace[\Detect\hat{S}\Detect]
        =
        \chi_{\Detect}(\hat{S})
        .
      \label{eq:DefCharacter}
      \end{equation}
      $\hat{S}$ is the representation of the stabilizer subgroup on the whole Hilbert space.
      The dimension of the representation is equal to the Hilbert space dimension.
      $\Detect\hat{S}\Detect$, on the other hand, is another one-dimensional representation
      of the same group, namely on the linear space proportional to $\ket{\PsiDet}$.
      The trace of a group representation is called the representation's {\em character} \cite{Weyl1950-0, Wigner1959-0}.
      And we have just shown that the phase factors are equal to the character of the stabilizer representation on the detection space.
      The situation discussed in the main text is when the stabilizer representation on the detection space is trivial.
      Then all phase factors are unity, and $\Detect\hat{S}\Detect = \Detect$.

      The representation $\Detect\hat{S}\Detect$ is one-dimensional and thus irreducible.
      Objects of the form \eqref{eq:appDefSymmProj} are called isotype projectors.
      They project onto the subspace that belongs to all isomorphic copies of the representation $\Detect\hat{S}\Detect$.
      This means that the detection state belongs to a certain (not necessarily trivial) irreducible representation.
      All ``symmetric'' states belong to the same representation or copies thereof.
      When the detection space becomes higher dimensional, i.e. when $\Detect = \dyad*{\psi_\text{d}^1} + \dyad*{\psi_\text{d}^2} + \hdots$
      and there are more than one detection state, the representation $\Detect\hat{S}\Detect$ is not necessarily irreducible anymore.
      The symmetry categorization of this situation is an important future project.

\end{document}